\documentclass[letterpaper,twocolumn,10pt]{article}
\usepackage{usenix2019_v3}

\pdfmapfile{+pdftex.map}

\usepackage{enumitem}
\usepackage{amssymb}
\usepackage{hyperref}

\usepackage{tikz}
\usepackage{amsmath}

\usepackage{filecontents}
\usepackage{tabularx}

\usepackage{booktabs}
\usepackage{tcolorbox}
\usepackage{multicol}
\usepackage{multirow}

\usepackage{listings}
\usepackage{xspace}

\usepackage{url}

\newcommand{\satelliteA}{Satellite A\xspace}
\newcommand{\satelliteB}{Satellite B\xspace}
\newcommand{\satelliteC}{Satellite C\xspace}

\newcounter{barrier}
\newenvironment{barrier}[2]
{
    \refstepcounter{barrier}%
    \label{barrier:#1}
    \begin{tcolorbox}
        \textbf{Barrier \thebarrier:} #2
    \end{tcolorbox}
}

\newcounter{strength}
\newenvironment{strength}[2]
{
    \refstepcounter{strength}%
    \label{strength:#1}
    \begin{tcolorbox}
        \textbf{Strength \thestrength:} #2
    \end{tcolorbox}
}

\newcounter{ofnote}
\newenvironment{ofnote}[2]
{
    \refstepcounter{ofnote}%
    \label{ofnote:#1}
    \begin{tcolorbox}
        \textbf{Of Note \theofnote:} #2
    \end{tcolorbox}
}

\newcommand*\dash{\unskip\kern.16667em---\penalty\exhyphenpenalty
        \hskip.16667em\relax
}

\makeatletter
\def\@listi{\leftmargin\leftmargini
    \parsep 1\p@ \@plus0\p@ \@minus\p@
    \topsep 2\p@   \@plus0\p@ \@minus\p@
    \itemsep1\p@ \@plus0\p@ \@minus\p@}
\let\@listI\@listi\@listi
\makeatother

\begin{document}

\date{}

\title{Unencrypted Flying Objects: \\ Security Lessons from University Small Satellite Developers and Their Code
}

\author{
{\rm Rachel McAmis}\\
University of Washington
\and
{\rm Gregor Haas}\\
University of Washington
\and{\rm Mattea Sim}\\
Indiana University \\
\and{\rm David Kohlbrenner}\\
University of Washington
\and
{\rm Tadayoshi Kohno}\\
University of Washington
} 

\maketitle

\begin{abstract}
Satellites face a multitude of security risks that set them apart from hardware on Earth. Small satellites may face additional challenges, as they are often developed on a budget and by amateur organizations or universities that do not consider security. We explore the security practices and preferences of small satellite teams, particularly university satellite teams, to understand what barriers exist to building satellites securely. We interviewed 8 university satellite club leaders across 4 clubs in the U.S. and perform a code audit of 3 of these clubs' code repositories. We find that security practices vary widely across teams, but all teams studied had vulnerabilities available to an unprivileged, ground-based attacker. Participants foresee many risks of unsecured small satellites and indicate security shortcomings in industry and government. Lastly, we identify a set of considerations for how to build future small satellites securely, in amateur organizations and beyond.
\end{abstract}

\section{Introduction}
\label{sec:intro}

Satellites have increased in popularity and magnitude in the past decade~\cite{satellites-increasing}. Satellites face a multitude of security challenges that differ from hardware on the ground today, such as the impossibility of physical access post-launch and limited contact periods as they orbit the earth. The decreasing cost of launching satellites~\cite{adilov2022analysis} has enabled more small players to launch, including amateurs and university teams. These amateur university teams typically operate smaller satellites, called ``smallsats.'' Academic smallsat use cases include testing university researchers' scientific payloads and giving students hands-on learning experiences. Despite being small projects and not-for-profit, university smallsats still face risks in space. One oft-cited risk is the Kessler Syndrome~\cite{kessler1978collision, kessler2010kessler}, where debris caused by satellites colliding with other objects could grow exponentially after reaching a particular threshold and thereby make future space exploration hazardous. Unsecured amateur satellites could harm the ability for emergency response given that amateur satellites can aide in this process~\cite{emergency-response}. Or smallsats might be used as ``kinetic weapons'', as discussed by interview participants and by Kurzrok et al.~\cite{kurzrok2018evaluating}. Smallsats could even incite international incidents or cause physical harm to humans, as we elaborate on in our results. Therefore, for the space ecosystem to be safe overall, we argue that amateur smallsats must also be secure. 

\textbf{Research Questions} To gain insights into computer security practices with smallsats, and with our focus on university satellite clubs as a window into the broader space, our first research question is:
\begin{itemize}
    \item RQ1: What are the security goals and practices of university satellite teams?
\end{itemize}

To understand whether and how to improve smallsat team security, we first need to understand existing goals and security practices. We analyzed this first through interviews with the leaders of 4 satellite clubs~\footnote{5 clubs met our search criteria; 1 club did not respond to inquiries for research, resulting in interviewing 4 clubs overall} (8 interviews in total, an average of 2 leader interviews per club) since club leaderships' perspectives on security affects the entire club's priorities. The aim was to surface concerns and practices across different teams rather than to make claims about generalizability. Thus, we chose qualitative methods and consistent analytical choices designed to summarize and describe our data. In other words, our research intends to uncover a richer and more nuanced understanding of the practices within this particular sample, rather than to make inferences about generalizability to other samples~\cite{Bertelsen2019}.

Secondly, we performed an audit over dozens of repositories from 3 of the teams we interviewed, parsing through code, design diagrams, and documentation. We focused on understanding the high-level security designs and choices of each club and club member, and comparing the designs in the code to their stated threat models in the interview; this type of design-level reconstruction and analysis on bespoke systems is not possible with automated tooling, therefore our analysis was manual. We detail our approach in Section~\ref{sec:code_results}.
To foreshadow our findings, security practices varied widely across teams, with none of the code we analyzed meeting all security measures needed to prevent malicious actors from sending commands up to the satellite while it is in orbit.

\begin{itemize}
    \item RQ2: What barriers to building satellites securely do smallsat teams face?
\end{itemize}
We identify a variety of barriers to building satellites securely. Club members  weigh tradeoffs between other goals (e.g., launching on time) and security. We also identify gaps in security knowledge, threat modeling, and implementation that serve as barriers to secure satellite implementations. For example, we find that security goals/non-goals vary greatly not only between clubs but \textit{among} club members, highlighting a non-uniform approach to security. Understanding these barriers can help us recommend solutions.
\begin{itemize}
    \item RQ3: What processes and tools can we recommend to improve smallsat team security?
\end{itemize}
Through a combination of interviews and our own knowledge as security researchers (4 of the authors), members of a university satellite club (2 of the 4), and as a leader within a satellite club (1 of the 4), we identify design recommendations and tools. These recommendations aim to bridge the gap between current satellite security practices and a more thorough security posture for smallsats appropriate to the university satellite context, i.e., groups with limited time, budget, and security expertise. We anticipate that some of these recommendations will apply to industry, as multiple participants indicated limited security considerations in industry. Additionally, many satellite club members end up working in the space industry (as we observed in our sample of leaders and our own satellite club experience), so understanding and addressing the security barriers at the university level may equip future space industry professionals with tools to build more secure industry space vehicles.

\textbf{Emerging Technology Security Research} It is well-known in the field of computer security research that computer security often lags technology innovations. For example, the modern automobile became computerized long before the automotive industry embraced computer security as a practice~\cite{koscher2010experimental}; likewise, the medical device industry incorporated computation and wireless communications before embracing computer security~\cite{halperin2008pacemakers}. The goal of our work is to contribute to the study of the emerging smallsat computer security domain --- a domain that, like automobiles and medical devices before, is exploding in innovation~\cite{satellites-increasing,adilov2022analysis} but, to our knowledge, has not yet ubiquitously embraced the need for strong and universal computer security practices. 

Whether today's smallsats are presently at risk of exploitation is speculative, but hacking enthusiasts have already demonstrated the ability to compromise government satellites~\cite{satellite-hack}, and commercial satellite communications networks have also been targeted~\cite{satcoms-hack-russia}. We return to assessing possible present and future risks in Section~\ref{sec:interview_results}. Studying university smallsats not only helps us explore the challenges and risks that these specific non-profit smallsats face, but given the proprietary nature of many smallsat manufacturers~\cite{willbold2023space, manulis2021cyber, falco2018vacuum} these open-source smallsat groups may be the closest possible window into satellite security overall.

\section{Background and Related Work}

\subsection{Satellite Applications and Overview} 
Though government-owned and commercial satellites are the most widespread today, hundreds of university satellites have launched since the 1980s, with launches increasing each decade~\cite{swartwout2018reliving}.
Even high school and middle school students have built and launched satellites~\cite{highshool-sat, middleschool-sat}.

University satellites support educational goals such as testing scientific payloads, hands-on learning experiences for students, and collecting space or Earth data. Some also serve as communication relays for amateur radio satellites (\textit{amsats}). 
Amsats facilitate ham radio communications, which can sometimes aid emergency response~\cite{emergency-response}. These emergency response utilities are another reason we believe that amateur smallsat security should be further studied. University teams often use amateur radio frequency bands for communication with their satellite. While ham radio usually requires communications to be unencrypted, it allows satellite command encryption~\cite{fcc_rule_97_211}.

Amsats are often supported by open-source infrastructure and volunteers who collect and relay telemetry using their own ground stations. One such example of this infrastructure is SatNOGS, an open-source ground station network~\cite{satnogs}. 
Multiple clubs we interviewed used SatNOGS to get data from their satellites from parts of the earth not accessible with their own ground stations. Many satellite clubs have open-source code repositories that are publicly accessible before and after their launch, 
which we discuss further in Sections~\ref{sec:interview_results}
 and~\ref{sec:code_results}. Another common resource is the the suite of communication protocols established by the Consultative Committee for Space Data Systems (CCSDS)~\cite{ccsdsOverview}. These protocol specifications cover a number of protocols across the network stack, some but not all of which include security measures in their specs.
 
\subsection{Satellite Architecture}
\begin{figure}[h]
\centerline{\includegraphics[width=0.4\textwidth]{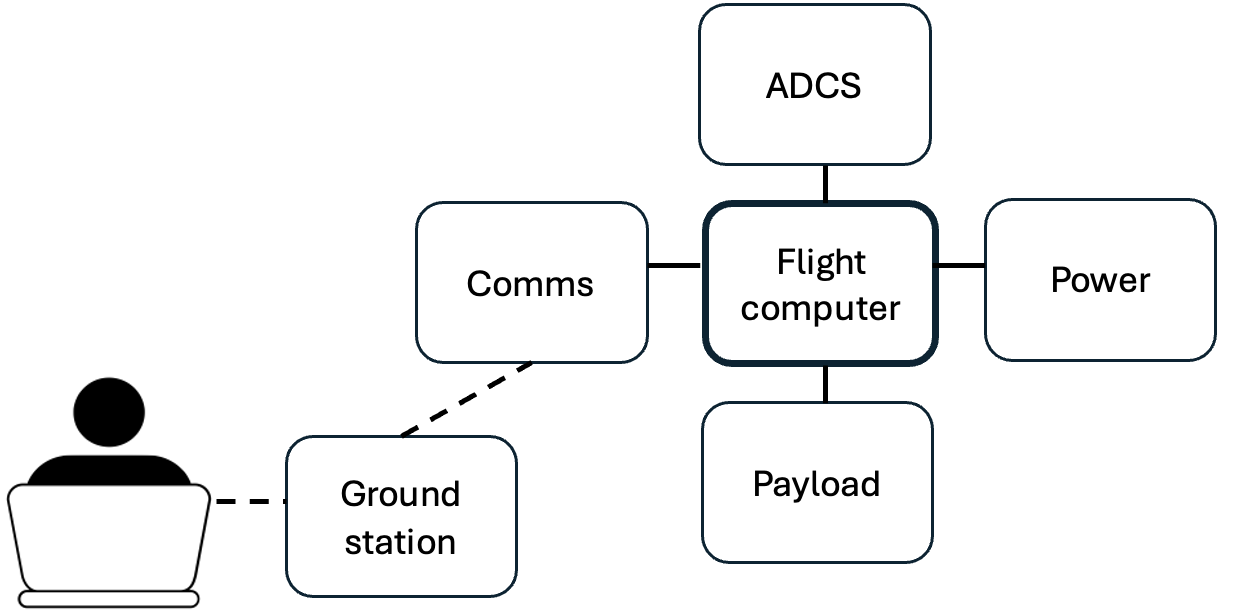}}
\caption{Overview of components in a generic smallsat. Ground station communicates with the satellite radio (comms). The flight computer communicates with all subcomponents, including power, ADCS, the satellite's payload, and comms.}
\label{fig:highlevelsatellite}
\end{figure}
At a high level, a smallsat consists of a few subcomponents, shown in Figure~\ref{fig:highlevelsatellite}:

\textbf{Ground station} The ground station is composed of a radio, antenna, and computer to send commands up to a satellite. A satellite might only receive commands from a single ground station, or a network of ground stations in different locations could help increase contact time with the satellite. The portion of communication sent from the ground station to the satellite is called the \textit{uplink}, and the inverse is called the \textit{downlink}.

\textbf{Communications space segment (Comms)} 
The comms segment on the satellite receives commands from the ground station via the uplink, and sends back telemetry (data about the satellite's state) and other command responses via the downlink. Some satellites have the ability to communicate directly with other satellites, especially if part of a fleet of satellites, but this is less common with university satellites.

\textbf{Flight computer} The flight computer schedules and facilitates communication between each subcomponent.

\textbf{Attitude determination and control system (ADCS)} ADCS helps control the orientation of the satellite, which is critical for keeping the satellite stable and ensuring that the satellite's solar panels are pointed toward the sun. Some satellites use propellant to maneuver, while others just use magnetorquers and reaction wheels.

\textbf{Payload} Payloads are the ``purpose'' of the  mission. Examples include cameras and novel batteries to test in space. 

\subsection{Satellite Security Challenges}
Securing satellites poses difficult security and reliability challenges, including:

\textbf{Contact constraints} In low Earth orbit, satellites take around 90 minutes to orbit Earth. Single ground stations can typically only contact a satellite for 7-10 of those minutes~\cite{devaraj2017dove}. And if the satellite were to enter an unrecoverable state, there is no way to physically access the satellite to reboot it.

\textbf{Latency constraints} Satellites may need to respond real-time for tasks like collision-avoidance or ADCS~\cite{jerosecuring}. Adding encryption and other security measures could increase the amount and size of packets needed when communicating. With the very limited bandwidth that many smallsats have, added packet size/count could lead to slower communications, ore even violate real-time constraints.

\textbf{Processing constraints} To meet their size and weight constraints for launch, smallsats typically have a limited power supply. Due to this, they have minimal processing power. Every processing step, including encryption and decryption, takes up valuable power and compute time~\cite{manulis2021cyber}.

\textbf{Custom Systems} Many satellites are largely bespoke systems, with custom protocols, operating systems, and hardware components. This makes it difficult to use existing security tools, such as fuzzing, as supported by Willbold et al.~\cite{willbold2024scaling}. Our security analysis in Section~\ref{sec:code_results} was therefore manual.

\subsection{Satellite Security Research}
A wide body of academic work has enumerated and taxonomized security and privacy threats of and to satellites~\cite{falco2021cubesat, manulis2021cyber, falco2018vacuum, cyr2023position, pavur2022building, mcamis2024over, jedermannrecord, koisserorbital, satpriv2}. Pavur et al. compiled a set of satellite threats enumerated in previous work and explained the history of satellite security incidents by decade~\cite{pavur2022building}. This threat modeling aided in our own threat modeling in Section~\ref{sec:code_results}.

While many papers have focused on threat enumeration, increasingly work has taken a more experimental approach~\cite{giuliari2021icarus, willbold2023space, jerosecuring, lane2017high, usman2020mitigating, jedermann2021orbit, ops-sat}. Willbold et al.~\cite{willbold2023space} analyzed firmware security of various satellites and surveyed industry members about security practices, highlighting a need for improved security among industry satellites. We build upon this work by more deeply exploring the university satellite security space, surfacing unique security barriers and needs.

Some work has highlighted the need for more security on university and other smallsats~\cite{kerlin2015small, kurzrok2018evaluating, saha2019ensuring}. Kurzrok et al. explored the theoretical extent that a hacked smallsat could travel under different propulsion methods~\cite{kurzrok2018evaluating}, showing the physical danger of unsecured ADCS telecommands. Kurzrok et al. also mentioned informal conversations among the smallsat community suggesting university teams often do not encrypt their telecommands and highlighted the need for a systematic analysis on university satellite security. Our work intends to fill this gap.

\section{Methodology}
\begin{table}
\caption{Interview participants by identifier. Security knowledge ranged from 1 (novice) to 5 (expert). Knowledge was based on self-reported scores except for D1. A * next to their score means that the respondent did not give a specific number, and therefore the score was an estimate based on their interview responses.}
\label{table:participants}
\centering
\begin{tabular}{l|c|c|c}
\toprule
    ID & Club & Industry & Security  \\
       & & Experience & Knowledge \\
    \midrule
    A1 & A & \checkmark & 3 \\
    A2 & A & $\times$  & 4 \\
    B1 & B & \checkmark  & 3 \\
    B2 & B & $\times$  & 3 \\
    C1 & C & \checkmark  & 2 \\
    D1 & D & $\times$  & 1* \\
    D2 & D & \checkmark & 2 \\
    D3 & D & \checkmark  & 5 \\
\end{tabular}
\end{table}
We used a combination of interviews and code analysis to explore the security practices of university smallsat teams. Our methodology involved five steps: 1) identifying university clubs online, 2) selecting clubs to interview and analyze based on publicly available code, 3) assessing the code to identify code-specific interview questions, 4) conducting interviews, and 5) analyzing the security of the code, cross-referenced with interview responses, based on our satellite-specific threat model.
We interviewed 8 participants across 4 clubs and looked through all publicly available repositories for 3 of those clubs.

\subsection{Finding Clubs to Analyze}
We used a search engine to identify as many university satellite clubs as possible, resulting in 41 overall. We then filtered for clubs with public code repositories and documentation. This resulted in 7 clubs, 5 of which are in the U.S. We chose public code repositories 1) because we did not wish to pressure clubs into revealing their code that could then lead to the possibility of them becoming de-anonymized, 2) avoiding conflict with their funders or institutions who might require that the code remain private, and 3) preventing violation of any ITAR guidelines, since many satellites have ITAR restrictions where certain implementation details cannot be shared to non-U.S. citizens. We chose not to interview members from the 2 clubs outside the U.S. because our research team is within the U.S. and regulations and policies may vary by country. Focusing on clubs within the U.S. allowed us to minimize variables including regulations and local educational cultures. 
Further, 1 of the 5 U.S.-based clubs did not respond to requests for interview and thus is not included in analyses. Finally, we interviewed Club D but excluded a code analysis after determining the available repositories corresponded to previous satellites our interviewees had not worked on (whereas more current code was not yet publicly available). We determined this only after Club D interviews started, so we still report their interview responses in Section~\ref{sec:interview_results}. While this may be a small set of clubs, there are not many clubs in the first place that have actually launched satellites per our searches, and a smaller set of those that have their code open source. 

\subsection{Interviews}
Our interviews take a standard approach by surveying developers to understand how security and privacy are treated in practice~\cite{wong2024comparing, 298888, ramulusecurity}. This approach allows us to understand gaps between research and practice and pose solutions towards reducing these gaps. We interviewed participants from 4 clubs.

\subsubsection{Recruitment and Screening}
To help answer RQ1, we especially recruited for members who worked on (and preferably led) satellite software. 
We also recruited satellite club leaders because their preferences likely affect satellite security implementation downstream. We found contact information through websites and public documentation. We reached out by email to each club and relevant club leaders, members, and alumni. Table~\ref{table:participants} summarizes our interview participants.

\subsubsection{Developing Interview Questions}
We constructed a set of interview questions based on our research questions. The first author constructed a semi-structured interview script, with pre-determined questions but allowing for flexibility (e.g., omitting questions, rewording, or asking follow-up questions as necessary depending on an interviewee's responses). This interview script was evaluated and revised by the authors and an external security expert. Two authors piloted the interview script with another author and an external colleague in robotics.

\subsubsection{Interview Procedure}
Interview questions below are labeled with their corresponding research questions. 

Two authors led the interviews virtually over Zoom meetings. The first author asked questions from the script while the second interviewer took notes, and both asked follow-up questions as necessary. We did not mention security until the security-specific questions to avoid biasing participants' responses about their goals and concerns when building a satellite~\cite{lazar2017research, redmiles2017summary, galesic2007sexual}. The interview script is included in Appendix~\ref{appendix:interview_script}.

\textbf{General Questions.}
To get a better understanding of participants' club experience, we asked about the satellites they worked on, what years they were in the club, and whether they had worked on satellites outside of the club.

\textbf{Mental Models.} We asked participants to draw a high-level diagram of how their satellites worked, and asked those with software experience to focus most on the software components. We used this diagram to help the participants and us as interviewers visualize further potential security threats. We did not yet mention security in this or the previous sections. We asked what goals they have when building satellites (Research Question 1) and what scenarios they were concerned about. These scenarios revealed whether security issues were a concern at the top of their mind (RQ1). We also asked about goals to further understand what tradeoffs to security they might be willing to make, which may serve as a barrier for implementing security (RQ2). Some of the wording and questions in this section and the following security-specific questions were inspired by Zeng et al.'s exploration of security and privacy perceptions of IoT users~\cite{zeng2017end}. 

\textbf{Security-specific Questions.}
We next asked whether they had any security-related concerns, or if they had other security concerns that they hadn't mentioned yet (RQ1, RQ2). We asked what they think the worst thing that could happen is if the satellite is not secure, who might attack a satellite, what discussions the club had about security, and how much they discussed security compared to reliability (RQ1, RQ2). These questions assessed participants' overall threat models by asking implicitly about assets (worst-case outcomes), adversaries (who might target the satellite), and vulnerabilities (what their security concerns were). We then asked what security tools they already used (RQ1) and what they wish existed (RQ3).

\textbf{Demographics.}
We asked about the participant's area of study, satellite team size and budget, their satellite industry experience, and their cybersecurity knowledge.

\textbf{Closing Questions.}
We asked if there was anything else the participants wanted to share and whether they expected us to ask anything else, a recommended practice in HCI research~\cite{lazar2017research}. We also asked several participants code-related questions relevant to their club's repositories to gain further insight into their procedures.

\subsubsection{Data Analysis}
For each interview transcript, both interviewers used emergent coding~\cite{lazar2017research} to note concepts that arose in each transcript. We compared themes that came up, then used axial coding to categorize these themes into groups. We translated these axial codes into a final codebook (Appendix~\ref{appendix:codebook}). For the frequency results in this paper, both interviewers coded based on the videos and the transcript. We coded these frequencies independently and calculated initial agreement using Cohen's Kappa, with substantial agreement between coders ($\kappa = 0.75$, $SE = 0.08$). We then further refined our codes and discussed differences in coding to reach agreement.

\subsection{Software Analysis}

Our analysis of the clubs' satellite software was split into two parts. First, two authors identified and analyzed repositories (along with public documentation and diagrams) belonging to each satellite club. As mentioned in the introduction, these analyses were entirely manual due to the goal of analyzing whether clubs implemented the high-level security goals that they stated in their interview, as well as the inherent difficulty in running automated tooling on disparate embedded platforms~\cite{abbasi2019challenges,willbold2024scaling}. From these analyses, we then derived high-level block diagrams (like Figure~\ref{fig:satelliteA}) of each satellite's architecture and system functionality. Second, four authors (all security and privacy experts) synthesized a threat model for each evaluated satellite based on these high-level diagrams. If our threat model differed from the threat model of the participants (as expressed in the interviews and corroborated by the software), we evaluated whether the code met the security requirements of \textit{our} threat model. We argue that this code analysis methodology serves as both a barometer on if the clubs' perceptions of their security postures are \textit{accurate} (i.e., the satellites behave how their clubs expect them to based on their mental models of the satellites, other interview responses, and comments in their code), as well as an indication of threat model \textit{suitability} (i.e., whether the threat models are appropriate for the satellite's goals, as judged by security and privacy experts).

Our manual code analysis was primarily guided by two authors' experience with satellite software. After identifying the clubs' Github organization and associated repositories, we filtered for security-relevant repositories. Since many of the repositories were for hardware designs or design collateral (Appendix~\ref{appendix:repositories:repo-stats}), the most relevant repositories tended to include software for ground stations, communication systems, and flight computers. For each of these repositories, we read the code at a high level to understand its intent and how it fits together with other subsystems. Rarely, we were able to additionally execute code that did not depend on bespoke embedded hardware (such as for \satelliteC). Based on these insights, we made diagrams for each satellite that describe (at a high level) how the security-critical components interact.

We then specified a general threat model (Section~\ref{sec:code:threatmodel}) to guide a more in-depth security analysis for each satellite. This threat model defines a remote adversary who can reverse-engineer the satellite's communication protocol and send arbitrary commands to the satellite. Our in-depth code analyses therefore typically began at each satellite's communication subsystem, specifically the handling function for received messages. We explored the threat surface of available commands, noting any reachable commands that might violate our participants' threat models, or confirm/deny any of their perceived risks or goals (Section~\ref{sec:worst-case}). If the participants noted that they implemented encryption or authentication, we verified whether this was the case. Finally, we also traced backwards from the participants' security and safety goals, and checked whether any internal state could lead to a violation of these.

\subsection{Scope and Limitations}
Our goal is to understand the computer security practices, barriers, and opportunities within the small yet growing community of university satellite clubs and their members. 
Our sample size was limited, in part because we recruited from a small and specific population (on the order of dozens of clubs based on our searches), which is itself a valuable endeavor~\cite{Bertelsen2019} and aligns with our broader research goals. Though our goal is not generalizability, qualitative methods can afford transferability~\cite{Daniel2019TACT}, such that lessons and recommendations uncovered here may be valuable in understanding other university satellite teams. Further, security vulnerabilities in only a few launched smallsats still have the capacity to cause harm to the entire space ecosystem, which is why it is so important to study this overlooked group of satellite operators. 
All participants interviewed were subsystem leads or heads of their clubs, so our results speak to club leaderships' perspectives on computer security and privacy. While there might be members within a club with different security knowledge than those that we interviewed, their existence does not detract from our findings with club leadership. Nevertheless, like any interview study, the results may suffer from self-reporting bias. We attempted to mitigate this through cross-referencing discussions of security with our manual code audit (except for Club D).

In terms of code analysis, it is possible that we missed relevant vulnerabilities given the scope of the repositories and the difficulty of using any automated tools on heterogenous, bespoke embedded systems. Because our security analysis focuses on the design-level, we attempted to mitigate this by anchoring our code analysis on the satellite-specific threat model we developed based on each payload, rather than analyzing for as many potential vulnerabilities as possible. 

\subsection{Our Perspective}
Our team of authors has experience with both attack-focused and defensive research. Two of our authors are members of a university satellite club and have experience developing components for a satellite. One author has experience as the lead of multiple subsystems within a satellite club. Three authors have expertise in participant-focused studies, including one author who is a social psychologist. Two authors have amateur radio communications experience. Multiple authors have embedded systems hardware and software expertise.

\section{Ethical Considerations}
Our institution's IRB approved the interview study under the exempt research category. Before each interview, we shared a consent form that participants agreed to verbally. Participants optionally consented to video recording through the Zoom recording functionality so that we could have an accurate script and record any drawn diagrams. All participants agreed to the recording option. One interviewer manually transcribed participant responses during each interview.
 
We paid each interview participant \$75 for their responses with an expected time of 1 hour per interview. This pay rate is higher than usual for interview studies; we wanted our pay to reflect an estimate of the time it would take out of an industry job to participate in our interview, since some of our participants were in industry. All questions were clarified to be optional. We anonymized study participants; the goal of this study is not to focus on individual clubs, but rather to understand what the range of security practices, preferences, and security barriers might be for university and other smallsat teams. To further protect participants, we sometimes obfuscate which anonymous participant said what in the combined results when we believe that it would not affect the interpretation of the response. For security vulnerabilities that we found that satellite clubs were not already aware of, we disclosed our findings to them. We offered our help to answer any questions or help resolve the vulnerabilities if needed.

\section{Interview Results}
\label{sec:interview_results}
We present below our interview results combined across clubs. Based on these results, we note strengths of security practices among teams, as well as specific barriers that might lead to difficulty implementing appropriate security measures. These results then inform our discussion of potential tools to help smallsat teams build more securely (Section~\ref{sec:design_recommendations}).

\subsection{Participants}
We interviewed 8 former and current university satellite club members about their satellite club experience across 4 clubs. See Table~\ref{table:participants} for a summary of participants. Participants consisted of club leaders, subsystem leads, software leads, or a combination of the above. 4 members were current and 4 were former members, graduating 2018 at earliest. 5 out of 8 members had experience in the space industry, both through internship and full-time roles. Security knowledge ranged from 1 (novice) to 5 (expert). Interviews lasted between 30 and 75 minutes.

\subsection{Satellite Goals}
To understand what tensions between security and other factors might exist, we asked participants about what overall goals they had when building their satellites. As a reminder, we did not mention security prior to the security-specific questions later in the interview. Answers fit broadly into the following categories: scientific missions, educating club members, educating community and other clubs, building the satellite cheaply, and building the satellite reliably. 

Clubs' scientific missions included taking images of Earth, testing battery systems in space, and relaying sensor data from remote regions. Clubs prioritized hands-on satellite development to educate club members, while outreach activities and payloads were often used to educate communities.  

Some clubs tried to build cheaply to stay within their own budget, while others also tried to build cheaply so that other clubs could replicate their designs in the future on a limited budget. They did this by relying minimally on off-the-shelf components, instead opting to build their own components from scratch or from open-source software and hardware. One such example was using the (unencrypted-by-default) OpenLST transceiver, a component with high manual cost to integrate but costing under \$100 per radio.

\barrier{budget}{Low-budget and education goals can lead to using homebrewed designs, which might influence the security tools used.}

\subsection{Difficulties faced}
\textbf{Non-technical difficulties were emphasized the most.} Participants discussed a wide range of difficulties within the club, both technical and non-technical. One of the most common categories of difficulties was non-technical issues relating to lack of various resources, with 7 participants bringing up difficulties in this category. Resource deficits included insufficient time for development, lack of people, limited experience, insufficient documentation within the club to transfer knowledge, insufficient public documentation to help clubs figure out what and how to build, difficulty finding materials and materials being discontinued, limited budget, and limited access to expensive physical testing infrastructure.

\textbf{Technical Difficulties Varied.} Technical difficulties mentioned varied by club and depended on both scientific missions and hardware used, though creating good software testing infrastructure and simulations was mentioned as a difficulty by 4 participants. Participant A2 provided a concrete example of this when they happened to find a memory corruption error: 
\begin{quote}While our attack surface was pretty limited, the worst case I suppose would be someone managing to gain remote code execution via a malicious uplink command entry point\ldots I discovered an array out of bounds issue at one point while testing myself when I noticed some data was corrupted and fixed it; we didn't have good automated tests to catch that kind of stuff even at the beginning. Most microcontroller code is written in C which is notorious for letting you shoot yourself in the foot with memory issues\ldots That combined with amateur software engineers is a potentially dangerous combo, especially for missions where a malicious actor could do more damage. (A2)
\end{quote}

\barrier{testing}{Insufficient testing --- potentially due to lack of time and other resources --- can lead to missing important satellite security flaws.}

\subsection{Security Discussions During Development}
The remaining portion of our interview focused on security. Only 1 participant mentioned insufficient security as a potential scenario of concern in our interviews before we explicitly asked about security concerns. 6 participants across 3 clubs said that security was discussed in the club once we asked. 

\textbf{Security perceptions differed \textit{across} and \textit{within} individual clubs.} 
Whereas some clubs did not authenticate or encrypt their commands, others did. However, the differences in security practices did not correspond with actual threats. For example, \satelliteB had a propulsion system but did not have encryption, while D3 said they would implement encryption even though all participants from Club D felt that there was minimal risk for their satellite.

There was also disagreement \textit{within} individual clubs about whether a security protection was employed (Club A, Club B) and whether security was discussed at all (Club D). We discuss this more in the code analysis section of individual clubs in Section~\ref{sec:code_results}. D3 said that reliability ``is mainly how I sold my [security] research and thesis.'' Their research was mainly on fuzzing, so this meant that finding security issues, such as memory corruption, was important for reliability as well. However the Club D leader said that there was no security discussion or security concerns with their satellite.

\barrier{inconsistent}{Inconsistent security goals and discussions within a club can lead to non-uniform implementation of security across the entire satellite system. One subsystem or code segment might have considered security while others have not.}

\textbf{Some clubs employed aspects of threat modeling in their design phase.} Both Club A participants said that security was discussed. A2 mentioned that their team met to whiteboard all the things that could go wrong. One design point they discussed in whiteboarding sessions was whether or not to allow software updates. On the plus side, being able to fix a bug after launching could save the mission. The downside is that this is complex and could introduce additional risk if incorrectly implemented. Thus, they decided not to include a software update mechanism. While software updates were discussed, the participant did not explicitly mention their value from a security perspective. Another example of at least partial threat modeling is participant D3's security fuzzing project; they mentioned using fuzzing to find bugs that would prevent the satellite being taken over and having firmware flashed onto it. 

\strength{modeling}{ Some clubs practice threat modeling in their satellite development process.}

\subsection{Satellite Threat Models}\label{sec:interview_threat}
\textbf{Participants discussed negative outcomes but mostly thought they were unlikely.} Participants brought up a wide array of potential threats to their satellite. 6 believed that these outcomes were unlikely, which often affected their security practices. For example, D1 believed that the satellite's sensor data was not sensitive and thus would not be a target.

We do not wish to imply that their risk assessments are misguided; based on our interview discussions, we have no evidence to suggest any club that has already launched their satellite was targeted by attackers. However, as threats change in the future and as the capabilities of university satellites change over time, these threat models may not hold. We see theories of this changing threat landscape, including space warfare, further into these results.

\textbf{Both club members and third parties had concerns.} For example, participants said that NASA was concerned that the lights on Club B's satellite could harm astronauts. The participants framed the blinding concerns as a safety issue rather than security issue.

B1 recalled that some outside groups, though they could not remember who, required Club B to think about security. B2 as club leader needed to ensure that all systems complied with International Traffic in Arms Regulations (ITAR~\cite{itar}), which sets some controls over how advanced various technologies on the satellite can be while still being open source, such as the propulsion and star tracking system. ITAR is enforced for national security and safety reasons rather than preventing the satellite itself from being exploited.
\ofnote{outside}{In some cases, security considerations might have been imposed from the outside, though the reasons may have been for safety.}

\textbf{Some threats were discussed during the design process while others were only discussed in the interviews.} Throughout our questions about security, club members mentioned a variety of possible ways to attack a satellite and potential security issues. These issues include sending up malicious commands, exploiting memory corruption vulnerabilities, performing remote code execution, denying service, unintentionally using malicious third-party code, and physical access pre-launch allowing someone to upload malicious code onto the satellite. However, some concerns only came up in the interviews and not while building the satellite. Remote code execution, third-party code, or physical access pre-launch were not discussed amongst club members while building the satellites. Some participants had industry experience where they were required to think more about security, which may have helped them with their threat modeling during the interview. The limited discussions of security within clubs despite many current members being aware of other potential security issues corroborates that security was not a main priority when building the satellites. We discuss what tools were used to mitigate these risks in Section~\ref{sec:security_tools}.

\barrier{notconsidered}{A lack of extensive threat modeling prior to launch means that some attack vectors were not considered.}

\textbf{Worst-case outcomes.} \label{sec:worst-case} For satellites that had propellant, one outcome that participants mentioned included sending malicious commands to ADCS to wreck their satellite, causing space debris. Even worse, B2 mentioned how the satellite might hit another satellite or the International Space Station (ISS). Some participants who worked on satellites without propellant or believed their propellant was not sufficiently powerful were not as concerned, with 1 participant mentioning that hackers could only retrieve public sensor data, and 2 others saying that they could lose communication with the satellite or otherwise be denied service. The above concerns were only considered during the interview, not prior to launch.

\textit{Physically harming astronauts} Club B had an extremely bright flashing light on their satellite. Club B participants mentioned blinding astronauts as a key concern in their design phase since their satellite was planned to launch from the ISS and would thus start out next to astronauts. Section~\ref{sec:satelliteA} shows what preventative measures they took to mitigate this risk. Club B participants mentioned that third-party providers who were helping launch the satellite were also concerned of this risk. This physical safety concern is the only threat considered prior to launch in this section of worst-case outcomes.

\textit{Violating Regulations} C1 mentioned that attackers could use satellites to violate regulation. For example, they could violate FCC guidelines and broadcast messages when they are not supposed to, or use a satellite's on-board camera to violate regulations and take images of other satellites rather than images of Earth~\cite{satelliteimagelaw}. 

\textit{Inciting International Incidents} Another concern that came up was that some foreign government could take hold of a university satellite from the U.S., use it to cause harm, and then blame it on the U.S. to limit U.S. presence in space, or to incite an incident. This concern came up without any prompting other than the interview script, and it shows an example of what the future of space warfare might look like:

\begin{quote} The worst case that I can think of is that a foreign adversary takes control of the satellite and [causes it] to run into the International Space Station\ldots [or] crash into some sort of satellite that's part of an international collaboration. At that point, the amount of paperwork that you're going to have would build a mountain high enough to get into orbit\ldots Any time that you have a satellite with a propulsion capability\ldots, you effectively have a kinetic weapon that can destroy other satellites\ldots A traditional adversary to the United States could take control of some U.S. satellite and say ``The student satellite is malfunctioning and it has crashed into our satellite \ldots This is a big international incident \ldots The United States can't control their satellites.'' (B2)
\end{quote}

This threat implies that any satellite with sufficient propulsion systems should consider its security, even if it's a university smallsat. Thus, teams should be aware about how this kind of threat might become more likely in the future as more warfare involves space~\cite{space-warfare}. 

\barrier{incomplete}{Incomplete threat models about why an adversary may target a club in the future (for example, to incite an international incident) can lead to differences between real threats and implemented threat mitigations.
}

\textbf{Potential Adversaries.} When we asked from whom the clubs' satellites might need protection, answers ranged from those hacking out of curiosity and boredom, to foreign government adversaries. Again, there was no indication that most of these adversaries or the adversaries below were considered prior to the interview. Some participants explicitly mentioned adversaries that they thought would not target their satellites, including foreign governments. Some explicitly mentioned intentions that adversaries would \textit{not} have towards university satellites, including not wanting money and that there would be no national security benefit to hacking them. 

\textit{Environmental Advocacy Groups} Some participants based potential adversaries off their experience. Club B received a letter from an environmental advocacy group against light pollution asking them to stop development of the satellite, and proposed that this group could be a potential adversary.

\textit{``Space Pirates''} Some participants posed that future space warfare might include ``satellite pirates'' or ``space pirates''. While this does have an existing definition regarding piracy of satellite-distributed content like satellite television, D2 had heard of satellite pirates as for-fun hackers who try to decode messages and use it against people. D3 heard of ``space pirates'' through previous work on a government-funded satellite security project. D3 defined space pirates as a satellite hacker that will become more important as space warfare becomes more prevalent~\cite{space-warfare}. D3 said that space pirates could even refer to someone with physical access to the satellite in the future: ``someone could have a spacecraft where they can approach your satellite or grab it out of space and modify its hardware.''

\barrier{expansive-threats}{Conducting an informed risk analysis of adversarial threats in space may be challenging, since discussion of some risks and predictions of risk, like ``space pirates'', require information not readily accessible in the open literature and therefore not as accessible to university satellite clubs.}

\subsection{Security Strategies and Tools Used} 
\label{sec:security_tools}
Clubs used a variety of tools and techniques, both effectively and ineffectively to try to implement security. In addition to whiteboarding out worst-case scenarios (Club A), participants said that they did manual security audits of code, are building fuzzers, and doing ``hardware debugging'' (Club D). As we will see in Section~\ref{sec:code_results}, some security protections employed were partially incorrect. 4 participants did not know what security tools were used or said that none were used. One tool for encryption mentioned was ``frequency modulation,'' so there was likely confusion about the term encryption. Based on participant ideas and our own analysis, we discuss potential future tools to help improve security in Section~\ref{sec:design_recommendations}.  Another security choice was to ``scrub commands'' from the public code (Club B). The command scrubbing in Club B was not effective for preventing adversaries from knowing the commands; see Section~\ref{sec:satelliteA} for more details.

\barrier{equipped}{Clubs may not be currently equipped with sufficient security tools and/or understanding of the security tools to effectively secure their satellite.}

\subsection{Security Tradeoffs}
All clubs made tradeoffs that directly impacted security, as any organization must do. Some clubs chose to forego confidentiality since they intended the communications downlink (satellite sending info to ground station) to be public for educational purposes, or were required for it to be public if they were within an amateur radio frequency band. One participant believed they could not encrypt the commands to the satellite in the uplink due to amateur radio laws, but this is not the case. Based on the authors' own experience in satellites clubs, this misconception is not uncommon.

\barrier{misunderstanding}{Misunderstanding in regulations might lead to implementing less security protections.}

Club A chose not to encrypt their uplink because they were concerned that they might implement it incorrectly and had seen other clubs ``shoot themselves in the foot'' when trying to implement encryption (A2). Club A decided it was a reasonable tradeoff because they did not think there was enough of a threat to justify encryption. Another participant said they cared a lot about maintaining simplicity of their software, which in turn would help with maintaining reliability (D2).

\barrier{reliability}{If a security solution has potential real or perceived negative reliability impacts, there may be resistance to the adoption of the security solution.}

\subsection{Security Issues in Industry}
\label{sec:industry_experience}
5 out of 8 participants had industry experience specific to satellites or in the space industry overall. 4 out of 5 only had industry experience after leaving the club, while 1 had industry experience during undergrad through internships. 4 participants said that industry takes security more seriously than their university satellite team (though one of these participants works now on rockets and not satellites). 2 participants said that there were insufficient security practices in industry. Some responses show a lack of priority in security in government satellites and possibly beyond:

\begin{quote}
    Any internships that I've applied for that have involved testing \ldots has been mainly on a reliability basis. (D3)
\end{quote}

\begin{quote}
    Not my specific company, but the space industry as a whole is hilariously bad at cybersecurity. It is comical and also terrifying, with the scope of projects in industry that have absolutely zero encryption. They're waking up to that now \ldots and it is being taken more seriously\ldots Cyber warfare has become much more of a concern in general (C1)
\end{quote}

D3 was going to be involved in developing the security of a government-funded satellite communications project that their university was involved in, until the funding for the security initiative was cut (at least for D3's institution):

\begin{quote}
    We were involved with [the satellite project] in the very early phases\ldots I think our group was chosen to do the security portion and then a month later when they were getting the cybersecurity budget sorted out for this project, they decided the budget would be \$0. (D3) 
\end{quote}

D3 worked on another satellite project outside of Club D:

\begin{quote}
    I was in integration testing on [a government satellite communications system]. As far as I could tell\ldots, they did no unit testing, no fuzzing, no security auditing. All the testing for catastrophic failures was done by people\ldots I was there for one or two potentially disastrous problems with the flight software that almost ended the project. I think most security stuff is left up to chance and most of these projects trust encryption as their only way to prevent hacks. (D3)
\end{quote}

\begin{tcolorbox}
{
\textbf{Of Note 2:} Security practices of satellites in industry and government may be insufficient relative to the satellites' capabilities.
}
\end{tcolorbox}

\section{Code Analysis}
\label{sec:code_results}

We now report on a code analysis of 3 clubs' satellite implementations, and compare and contrast these with security perceptions from our interviews. This code analysis encompasses dozens of repositories and many thousands of lines of code; Appendix~\ref{appendix:repositories} lists statistics about these repositories.

\subsection{Threat model}\label{sec:code:threatmodel}
To inform our code analysis and satellite-specific threat models, we first establish a more general threat model. We assume an adversary on the ground who only has access to public documentation about the satellite, i.e., is not an insider to the club. Public documentation includes all open-source code repositories, documentation within those repositories, or any other information about the satellite available on the internet. 
Public documentation also includes network traffic, since adversaries could reconstruct the protocol based on observing commands sent from the ground station~\cite{borisov2007generic}. The adversary can construct their own ground station or use a ground station as a service company and send commands up with a software-defined radio or other tool.

We intentionally exclude pre-launch physical access to satellites and their underlying hardware by the adversaries from our analysis; therefore, supply-chain attacks and physical access are not part of our threat models. We are \emph{not} saying that these two threats could not compromise the operation of a university satellite. Rather, given that building satellites to handle supply chain adversaries is an emerging area of research~\cite{jerosecuring}, we do not expect (and did not see) any defenses for these threats in the code that we analyze, and hence we define it as out of scope in our analysis. Second, all satellites analyzed were vulnerable to a much less privileged remote adversary, making whether the satellite could defend against supply chain attacks not relevant. To clarify, excluding these results does not impact our interview results, as the participants were not told to consider a certain threat model. Still, we explicitly acknowledge that future works should consider supply-chain attacks, especially as satellites become more ubiquitous and capable.

\subsection{\satelliteA}
\label{sec:satelliteA}
\satelliteA launched in 2018 and decayed from orbit in 2020. \satelliteA had two main science missions: 1) to test a battery technology in space and 2) to implement a ``flashsat'' using a 40000 lumen LED panel, designed to be seen from space by the naked eye. For the rest of this section, we use the term ``flash'' to refer to the LED panel, rather than persistent storage or the process of installing software updates into that storage.

We find that the \satelliteA designers were primarily focused on mitigating flash panel safety concerns, which they did successfully via precise state transitions (Figure~\ref{fig:satelliteA}). However, we identify availability and confidentiality issues.

\subsubsection{System Overview}

\textbf{Bootloader}
The command/control/communications (C3, aka flight computer) software component consists of a low-level bootloader and the main real-time OS (RTOS). 
The bootloader is a tiny program which runs when the C3 first starts up. Its main task is to interface with an external MRAM (\textit{m}agnetoresistive \textit{r}andom \textit{a}ccess \textit{m}emory) chip to load the larger operating system image. MRAM is inherently tolerant of radiation (i.e., not prone to bit flips), and is therefore a safe place to store the large, mission-critical operating system.

\begin{figure}
    \centering
    \includegraphics[width=0.7\linewidth]{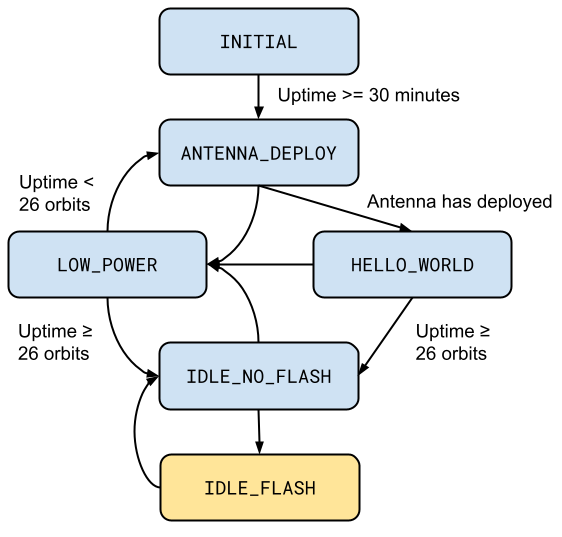}
    \caption{\satelliteA state transition diagram. The satellite can only flash during the \texttt{IDLE\_FLASH} state, which is only reachable after a wait period of 26 orbits for safety reasons. State names correspond closely with satellite actions taken during those states, except for \texttt{HELLO\_WORLD} which acts simply as an idle state. The \texttt{LOW\_POWER} state is activated any time the satellite detects that the battery levels are too low.}
    \label{fig:satelliteA}
\end{figure}

\textbf{Operating System}
The main OS is built on FreeRTOS 9.0.0, a widely-used RTOS with a broad range of hardware support. The OS performs low-level hardware initialization, bootstraps the scheduler and other core subsystems, and finally launches RTOS tasks to manage individual hardware components. These tasks include battery charging (\texttt{BATT}), antenna deployment (\texttt{ANT}), and (of note) flash panel activation (\texttt{FLASH}), among others. The RTOS schedules these tasks based on the satellite's current state as shown in Figure~\ref{fig:satelliteA}. 

\textbf{Ground station} \satelliteA used two distinct ground station networks to collect downlink telemetry for portions of orbit not reachable by the club's own ground stations. At first, the club solicited amateur radio observations by offering an online upload portal for observations. Later, the club began using a new local decoding feature in the SatNOGS open source ground station network, which required them to have their downlink protocol be publicly accessible. Based on comments in the code and interviews, the club thought of these ham radio operators as honest-but-curious. 

\subsubsection{Threat Model Comparison}
\noindent
\textbf{Participant Threat Models} As reported in Section~\ref{sec:interview_threat}, \satelliteA designers and NASA had safety concerns related to the flash panel. \satelliteA's designers also had \textit{confidentiality} concerns with the satellite's comms protocol, mentioning that they ``scrubbed commands'' from their public repositories as an ``operational security measure.'' The interviewees discussed various potential adversaries, including an environmental advocacy group who sent them a letter prior to launch and honest-but-curious ham radio operators. Another threat that A2 mentioned was sending commands in such a way that their battery system payload would explode.

\textbf{Does Participant Threat Model and Code Match?} \satelliteA's designers successfully mitigated the astronaut-blinding threat by ensuring that the satellite is far enough away from its launch vehicle and the ISS before it can begin flashing. We show the implementation of this mitigation in Figure~\ref{fig:satelliteA}. We find evidence in the code that the authors thought about the safety of every state transition. The satellite's ``flash immediately'' command, for example, does not activate the flash panel unless the satellite is in the \texttt{IDLE\_FLASH} state. 

However, the club's confidentiality expectations were not met. All commands were still available in the code, and all information necessary to build \satelliteA's ground station was publicly available during its lifetime. Even if commands were scrubbed, a motivated adversary could reconstruct the protocol based on their unencrypted uplink network traffic. 
Since the satellite commands are not authenticated, adversaries could have interacted with the satellite, although neither we nor the participants saw evidence that this occurred. We were unable to verify whether sending up a certain set of commands could cause the satellite to explode.

\textbf{Does our Threat Model Match?}
We also believe the most significant risk is physical harm to astronauts, and this outcome is successfully mitigated by precise state transitions. While we do not evaluate the feasibility of the exploding satellite attack, we do think that there are other ways to deny service not discussed by participants. Available commands included simple pinging, rebooting the satellite, enabling/disabling radio responses (presumably to comply with amateur radio laws), enabling/disabling the flash panel, and sending a command to flash the panel immediately. An attacker with a ground station could arbitrarily disable both radio responses and the flash panel, as well as reboot the satellite. If the club does not think to re-enable radio responses or the flash panel they may believe that the satellite has failed in some more severe way, affecting the satellite's \textit{perceived availability}.

Before the satellite decayed, we count several reported CVEs~\footnote{\url{https://cve.mitre.org/cgi-bin/cvekey.cgi?keyword=freertos}} for FreeRTOS 9.0.0. While we did not evaluate whether these CVEs were applicable for \satelliteA's configuration, dependence on external code that is not updated regularly is a well-known security risk. \satelliteA also lacks a firmware update mechanism, which increases this risk. 

\subsection{\satelliteB}

\satelliteB launched in 2020 and decayed in 2023. This satellite tested a propellant system and a second payload that we omit since it is not relevant to our study and revealing it could facilitate de-anonymization of the club.

\satelliteB's security posture heavily relies on \textit{obfuscation}. While the satellite's source code was public during its lifetime, B2 mentioned that extensive protocol layering made it ``pretty impossible to send commands.'' Additionally, the radio implementation (which was provided to the club by amateur radio collaborators) was not public for our analysis. Focusing instead on the satellite's bus architecture, we find that the radio broadcasts packets directly on a CAN bus for consumption by subsystems. These packets are implicitly trusted by all subsystems, indicating that the radio was the main security boundary for the satellite. 

\subsubsection{System Overview}
\satelliteB was constructed out of many individual microcontrollers implementing various subsystems, such as ADCS and propellant control, and connected via a shared CAN bus. The radio system was provided by AMSAT  (the Radio \textbf{Am}ateur \textbf{Sat}ellite Corporation~\cite{amsat}), and was also connected to the satellite's CAN bus; received uplink packets were written directly to CAN and filtered/interpreted by the subsystems, while telemetry packets were downlinked as they were generated by the microcontrollers. The satellite used an open-source high-level operations tool to issue commands that get translated to lower-level messages on the CAN bus.

\subsubsection{Threat Model Comparison}

\noindent
\textbf{Participant Threat Models} 
B1 was in Club B while their satellite launched, while B2 only started after the launch. Both mentioned concern about the thrusters and propellant on-board. B1 did not think that the thruster was strong enough to be used as a projectile. Instead, they were concerned with denial-of-service through the attacker sending up many messages to the satellite, either to flood the comms system or to send commands to use and drain the propellant. B1 thought that the adversary would be a ``for-fun hacker.''

B2 on the other hand \textit{does} believe that the propellant could have been a problem, and that the worst case was inciting an international incident by either moving the satellite to cause debris or to knock into space vehicles owned by other countries (See Section~\ref{sec:worst-case}). Because of this threat, B2 thought an adversary might be a foreign nation.

\textbf{Does Participant Threat Model and Code Match?} \satelliteB's architecture implicitly places all security expectations onto the radio. The satellite's internal bus is a broadcast CAN bus where every node can see every message, but filters only for messages it can handle. The radio writes all received packets to the CAN bus, presumably after unwrapping some link-layer framing information. As such, the AMSAT radio would have been our primary focus for code analysis had it been made public. We instead checked  AMSAT's documentation for this line of systems and \satelliteB's open-source ground station, and did not see evidence of security mechanisms in either. Our participants did not recall there being comms link encryption, further supporting this assumption.

To meet Club B's threat models for propulsion system safety, we believe that command authentication would be necessary to prevent either denial-of-service or an international incident, as envisioned by our interviewees. A2 said that radio sequence numbers could make it harder for adversaries to ``patch in a command,'' but we did not find evidence of sequence numbers being used in either AMSAT's public documentation or \satelliteB's ground station. Therefore, there is a mismatch of safety expectations between \satelliteB's designers and their code.

Based on interview responses, we believe \satelliteB's complex and layered comms protocol led club members to mistakenly believe it was \textit{confidential}; one participant said it was ``encrypted by obfuscation'' but ``not industrial-grade encryption.'' However, their perceived threats did not necessarily require confidentiality to be successfully mitigated.

\textbf{Does Our Threat Model Match?}
We find evidence that the \satelliteB repositories were public during its deployment. Since we found no authentication or encryption of commands, we argue that a motivated nearby adversary could have discovered which commands exist and how to send them.

We agree with our interviewees that denial of service and causing an international incident are the worst-case scenarios. We find commands primarily for changing the ADCS magnetorquer settings, powering on/off various subsystems, and pulsing the propulsion system. An adversary on a limited budget (Section~\ref{sec:code:threatmodel}) could use these commands to harm availability, e.g. by tumbling the satellite, turning off subcomponents, or exhausting propellant. 

\subsection{\satelliteC}

Club C implemented a complete open-source amateur satellite system, used so far in two launched satellites.
The most recent satellite was launched in 2024.

We find that the main security boundary for \satelliteC is the comms subsystem, and that \textit{authentication} (rather than \textit{obscurity}) is the primary security mechanism. We identify two small implementation errors in their authenticity mechanisms and a lack of defense-in-depth. Beyond the comms subsystem, we find advanced functionality enabled by \satelliteC's Linux-capable C3. Unique among the satellites we analyzed is a software update capability. 

\subsubsection{System Overview} 

\satelliteC's architecture consists of a variety of interconnected ARM-based processing units, ranging from microcontrollers running an embedded RTOS to Linux-capable CPUs. These processors are all attached to their relevant subsystems (ranging from ADCS to solar panels) and a central CAN bus for intercomponent communication. The C3 is built as a standard Python application running on Linux, enabling advanced functionality such as software updates.

\satelliteC's payloads are a novel high-speed comms link and a camera. As part of the club's outreach goals, they want to take images with the camera from space and then transmit them to mobile ground stations built by local high schools. 

\textbf{Communications} 
\satelliteC's comms protocol follows the official CCSDS protocol of the Unified Space Link Protocol (USLP). They use a keyed HMAC built on 256-bit SHA3 for authenticity. Sequence numbers are used to attempt to prevent ``repeat attacks,'' (we believe they meant \textit{replay} attacks) as noted in their public documentation.
Their high-speed comms system, operated by the C3, is separate from the main comms link. Commands for taking photos and quickly transmitting them first go through the main C3 computer (following the packet structure), then are immediately routed to and handled by the high-speed subsystem.

\subsubsection{Threat Model Comparison} 
\noindent 
\textbf{Participant Threat Models} C1 identifies both confidentiality and regulatory concerns, stating they did the ``bare-minimum of encryption and due diligence''. They also expressed concerns related to the camera~\cite{satelliteimagelaw} violating NOAA regulations by taking images of space (and orbiting objects) rather than Earth. Command integrity is required to prevent the risk of an adversary sending a malicious camera command.

\textbf{Does Participant Threat Model and Code Match?} 
C1's stated threat model is partially addressed in the code. Command integrity concerns are addressed in the comms uplink aside from two implementation issues. We saw no evidence of uplink encryption, contradicting C1. However, based on their concern about violating regulations, uplink confidentiality is unnecessary; only uplink \textit{authenticity} is needed.

We find that \satelliteC developers consider the comms subsystem to be the satellite's main security boundary. Each message includes both an HMAC and a sequence number to prevent replay attacks, though there were two slight errors. 

First, we found an off-by-one error on packet sequence numbers, allowing for replay of the last received packet an arbitrary number of times until another authentic command increments the internal sequence number.  We confirmed that this replay attack works by running the C3 software locally and successfully sending the same packet multiple times. 

We argue that sequence numbers only protect commands which change the system state independently of the number of times they are invoked (\textit{idempotent} commands). Our code analysis revealed a rich set of \textit{non}idempotent commands, ranging from component- and system-level resets to real-time clock updates. 
We did not fully investigate the idempotency of every command (since there are very many), but did check some of the more sensitive ones related to availability. For example, we verified that the magnetorquers are configured via ``setpoint'' commands that indicate the desired torque level along each axis, rather than specifying \textit{changes} to these levels; this makes the magnetorquer commands idempotent and mitigates against the effects of a potential replay attack.

The other issue is a timing vulnerability on the packet HMAC check due to a non-constant-time comparison operator. An attacker with sufficiently precise timing information could possibly deduce the expected HMAC for an arbitrary packet. We did not check the feasibility of this attack since there are some major limitations, such as the HMAC's 32-byte length and a lack of prior work on ground-to-space timing side channels. Nevertheless, we reserve establishing practical exploitability for future work.

Additionally, we did not find evidence of specific security measures that were taken in the rest of \satelliteC's code, indicating that defense-in-depth was not a concern for the developers (which also matches C1's threat model). For example, we found that no integrity or authenticity checking is done on update files beyond the HMAC/sequence number checks as update packets are received. Therefore, if an attacker were able to bypass these checks they may be able to establish persistence on \satelliteC via a malicious software update. Similarly, we find no specific security measures taken to prevent illicit imaging of other satellites. Instead, the camera is simply triggered by a command from the C3 when a corresponding radio command is authenticated.

\textbf{Does our Threat Model Match?} We agree that regulatory concerns in space are important to consider. Like other satellites, another potential threat is denial of service. The replay attack vulnerability could harm satellite availability.

\section{Security Recommendations for Smallsats} 
\label{sec:design_recommendations}
Informed by our interviews, code analysis, and our own satellite club experience, we discuss approaches that could help improve security practices among university and smallsat teams. Some recommendations apply to other amateur satellites and even industry, as our participants and prior work~\cite{willbold2023space} show industry security also needs enhancement.

\subsection{Open-Source Tools and Techniques} 
\textbf{Software/Hardware} Any tool should consider the incentives impacting the satellite development process, including the particular emphasis on satellite systems' reliability (Barrier~\ref{barrier:reliability}). Any proposed smallsat security tool would ideally have reliability benefits. 
An example of an effective security and reliability tool is updating the hardware of a commonly used smallsat component. 
We found that multiple university clubs rely on the open-source OpenLST transceiver package\footnote{\url{https://github.com/OpenLST}} for their communications. One comment in OpenLST code is ``TO-DO: handle encryption''. In their documentation they say that they do not implement encryption. 
A future project could be to fork this repo and update it with encryption and authentication. Since OpenLST is on longer maintained, hardware updates alongside security improvements would increase reliability and thus likelihood of adoption.

As university clubs prioritize learning, a fully-furnished secure smallsat framework may have limited appeal. However, an out-of-the-box cryptographic comms solution could be valuable, as comms systems often define security boundaries. Authentication-only tools may also be useful as confidentiality is not applicable to every club's threat model.

\textbf{Checklists} 
Security considerations in smallsat repositories are incomplete. B2 suggested a security checklist. This could include recommendations for cryptographic libraries, secure reference code, design guides, and organizational security practices. A2 raised concerns about cryptographic techniques affecting reliability, so the checklist could integrate lessons from embedded software and network security.

\subsection{Operational}
\textbf{Actions Within Club} We saw partial threat modeling in Club A and Club C. Satellite clubs could incorporate more thorough threat modeling into the design stage. Clubs could use tools such as threat modeling cards~\cite{mead2018hybrid, securitycards}. Though suggesting tools does not diminish the time constraints clubs face.

\textbf{Third-party Actions} To address these competing incentives, a trusted third party, such as NASA, could recommend or require security measures before launch. NASA's Cubesat Launch Initiative Program~\cite{nasa-csli} would be a particularly good location to promote good security practices in educational and non-profit satellites, since they are involved in the process of many universities' smallsat development. NASA could provide some information about the importance of smallsat security in their educational resources, such as their Cubesat 101 book~\cite{cubesat101}, or even incorporate the secure design checklist suggested above. 
For those not involved in NASA CSLI, launch providers could also set some requirements for satellites. Trusted third parties could also offer threat modeling support or serve as security contacts for university teams.

A ``No Encryption, No Fly Rule'' was proposed by authors based on their analysis of the damage propulsive systems on smallsats could cause~\cite{kurzrok2018evaluating}. We propose expanding this rule to cover integrity and authentication, since these are even more important than confidentiality for many amateur smallsats.

\section{Conclusion}
We examined the security practices and perspectives of university smallsat teams through interviews and code analysis. While teams implemented some security measures, they were neither systematic nor complete. We observed impressive engineering efforts and believe improvements in tools, organizations, and third parties involved in launch can improve smallsat security. These changes will hopefully align with teams’ goals of reliability and education while mitigating risks to the broader space community.

\section*{Acknowledgments}

This work was supported in part by the University of Washington Tech Policy Lab and the National Science Foundation under award CNS-2207019.



\bibliographystyle{IEEEtran}
\bibliography{references}

\appendix
\section{Interview Protocol}
\label{appendix:interview_script}
\textbf{Introduction}
Good morning/afternoon. Our names are [Researcher 1] and [Researcher 2] and we’re working with [Institution 1] and [Institution 2] on this study. Thank you for coming. You are one respondent in our sample who has been asked about their university satellite club experience. We’re doing this so that we can understand the challenges and complexities of building university satellites, and our ultimate goal is to have these interviews help us come up with and build future open-source tools for small satellites.\\\\
\textit{[Other introduction, recording consent.]} \\\\

\textbf{Experience/Background}
\begin{itemize}
    \item What satellite clubs are you a member of or have you been a member of? What years?
    \item What role or roles do/did you perform at the Satellite Club? For example, electrical, structures, admin, communications, etc.
    \item How many years of experience do you have working on satellite-related projects?
    \item Which satellites have you worked on developing? You can count ones that are still in progress and have not gone into space yet.
\end{itemize}
\textbf{Mental Models}
\begin{itemize}
    \item For this next part, I’d like you to pick one satellite you worked on if you worked on multiple. I’d like you to spend 3-5 minutes drawing a diagram of how all of the subcomponents of the satellite are connected together, including the ground station. I will open up a whiteboard on Zoom. If you don’t want to do a Zoom whiteboard, you can draw on your preferred format instead. This doesn’t have to be perfect. 
    \item Could you walk us through the diagram?
    \item What goals do you have when building your satellite?
    \item What are the most difficult aspects of building your satellite?
    \item \textit{If they haven’t already talked about engineering/technical details in the previous question:} 	What are the most difficult technical aspects of building a satellite?
    \item Are there any scenarios you’re concerned about when building and/or launching a satellite? \\
    \textit{If yes: }
    \begin{itemize}
        \item How do you address these concerns? 
        \item \textit{If satellite was launched: } Did any of those scenarios end up happening?
        \item \textit{If scenarios did end up happening: } If you were to advise your past self about preparing for different scenarios, what advice would you give? \\
    \end{itemize}
\end{itemize}
\textbf{Security-Specific Questions} \\
\begin{itemize}
    \item \textit{If they have not brought up security:} One type of concern we’re interested in is cybersecurity concerns. From this point on, when we use the word secure or security, we mean in terms of cybersecurity. Do you or did you have any concerns related to security about your satellites? You might not have any such concerns -- that’s fine, and we’d like to hear about that too. 
    \item  \textit{If they brought up security organically:} From this point on, when we use the word secure or security, we mean in terms of cybersecurity. Do you or did you have any other security concerns that you haven’t mentioned yet?
    \item What do you think is the worst thing that could happen if the satellite is not secure?
    \item \textit{If adversaries not already discussed:} From whom do you think satellites might need protection in the worst case?
    \item Satellite teams have many reasons to not focus on security. Was security discussed? If so, how much? 
    \begin{itemize}
        \item What were/are the conversations often about?
        \item \textit{If adversaries were not already discussed:}
	    Did you discuss what entities your satellites might need protection from?
    \end{itemize}
    \item How much was security discussed compared to reliability?
    \item What tools or techniques, if any, does or did your team already use to help with security?
    \item \textit{If satellite groups have tried to make their satellites secure:} Have you experienced any challenges in trying to incorporate security into your satellite?
    \item \textit{If technicaly challenges not mentioned in previous question:} What technical challenges did you have trying to incorporate security into the design or implementation stage?
    \item Do you think there are tools or practices your team could incorporate to help improve security for future university satellites? These tools do not need to exist yet. The answer to this question can also be no.
    \item How confident are you that you are able to build a satellite that is secure on a scale of 1 to 5, with 5 being the most confident? Why did you rate yourself that number?
    \item How would you rate your knowledge of computer security on a scale of 1 to 5, with 1 being a novice and 5 being an expert? Why did you rate yourself that number?
\end{itemize}
\textbf{Demographics}
\begin{itemize}
    \item What is/was the size of your satellite team?
    \item What is/was the team’s budget?
    \item What is your major / what was your major in college?
    \item Have you taken a security course or learned about cybersecurity? Before or after you worked on a satellite?
    \item Have you worked on satellites in industry?
    \begin{itemize}
        \item \textit{If yes:} What differences, if any, did you notice in the security practices and tools between university and industry satellites?
        \item \textit{If there were differences:} Based on your experience, if you were to work in your satellite club again, is there anything from a security standpoint that you would do differently? \\
    \end{itemize}
\end{itemize}
\textbf{Closing Questions}
\begin{itemize}
    \item \textit{(Ask club-specific questions)}
    \item Are there any other questions you expected us to ask?
    \item Is there anything else you want to tell us about your satellite-building experience?
\end{itemize}

\setlistdepth{9}

\setlist[itemize,1]{label=$\bullet$}

\setlist[itemize,2]{label=$\bullet$}

\setlist[itemize,3]{label=$\bullet$}

\setlist[itemize,4]{label=$\bullet$}

\setlist[itemize,5]{label=$\bullet$}

\setlist[itemize,6]{label=$\bullet$}

\setlist[itemize,7]{label=$\bullet$}

\setlist[itemize,8]{label=$\bullet$}

\setlist[itemize,9]{label=$\bullet$}

\renewlist{itemize}{itemize}{9}

\section{Interview Response Codebook}
\label{appendix:codebook}
\begin{itemize}
    \item Goals when building satellite
    \begin{itemize}
    \item Reliability
    \item Scientific missions
    \item Education of club members
    \item Education/Outreach outside of club
    \item Building cheaply
    \end{itemize}
    \item Difficulties
    \begin{itemize}
        \item Non-technical
        \begin{itemize}
            \item Safety/harm to themselves
            \item Access to resources
            \begin{itemize}
                \item Time
                \item People
                \item Experience
                \item Communication/Knowledge transfer
                \item Money
                \item Existing external documentation
                \item Materials
                \item Physical Testing
            \end{itemize}
            \item Technical
            \begin{itemize}
                \item Specific subsystem implementation
                \item Integrating subsystems together
                \item Building from scratch
                \item Reliability
                \item Getting design stage correct
                \item Minimizing complexity
                \item Testing
            \end{itemize}
        \end{itemize}
    \end{itemize}
    \item Security
    \begin{itemize}
        \item Was security discussed in the interview before we explicitly asked?
        \begin{itemize}
            \item Yes
            \item No
        \end{itemize}
        \item Was security discussed within the club?
        \begin{itemize}
            \item Yes
            \item No
        \end{itemize}
        \item Respondent perceived threat likelihood 
        \begin{itemize}
            \item ``Unlikely''
        \end{itemize}
        \item Who was concerned
        \begin{itemize}
            \item Third-party concerns (launch provider, NASA)
            \item Within club
        \end{itemize}
        \item When did security concerns come up?
        \begin{itemize}
            \item Security concerns discussed in club
            \item Security concerns only came up when we asked in interview
        \end{itemize}
        \item Security Issues
        \begin{itemize}
            \item Denial of service
            \item Malicious commands
            \item Memory corruption
            \item Remote code execution
            \item Physical access pre-launch
            \item Malicious third-party code
        \end{itemize}
        \item Security Impact/Outcomes
        \begin{itemize}
            \item Wrecking own satellite
            \item Harming other satellites
            \item Propelling the satellites
            \item Blinding astronauts
            \item Hitting International Space Station (ISS)
            \item Violating regulation
            \item International incident
            \item Crash causing debris
        \end{itemize}
        \item Adversary
        \begin{itemize}
            \item ``Hacker''
            \item ``Satellite pirates''
            \item Environmental advocacy group (specific group anonymized)
            \item Foreign adversary
            \item Amateur radio operator
            \item Personal vendetta
            \item For-fun hackers (curiosity, boredom)
        \end{itemize}
        \item Non-Adversary (when people specifically mentioned adversaries they were not concerned about)
        \begin{itemize}
            \item Foreign adversary
        \end{itemize}
        \item Security choices
        \begin{itemize}
            \item No encryption
            \item Protections other than encryption
            \item Exploring worst-case scenarios
            \item Encrypting uplink
            \item Obfuscation
            \begin{itemize}
                \item Of code (not public)
                \item Of commands
            \end{itemize}
        \end{itemize}
        \item Security Tradeoffs
        \begin{itemize}
            \item Intended public downlink
            \item Open-source implementation
            \item Maintaining simplicity
            \item Maintaining functionality
        \end{itemize}
        \item Challenges Incorporating Security
        \begin{itemize}
            \item None/not sure
            \item Misunderstanding of regulation (e.g., thinking you cannot encrypt amateur radio telecommands to satellites)
            \item Monitoring code across systems/people
            \item Time/effort
            \item Expertise
        \end{itemize}
        \item Security Tools Used
        \begin{itemize}
            \item None/not sure
            \item Fuzzers
            \item Manual auditing
            \item ``Hardware debugging''
            \item Encryption
            \item ``Frequency modulation''
            \item Obfuscation
        \end{itemize}
        \item Industry Security Practice Differences
        \begin{itemize}
            \item Equal practices between university and industry
            \item Industry more secure
            \item Lack of good security practices in industry
            \item No satellite industry experience
        \end{itemize}
        \item Future Security Ideas
        \begin{itemize}
            \item None/not sure
            \item Non-technical
            \begin{itemize}
                \item Check organizational security (e.g., document access)
            \end{itemize}
            \item Technical
            \begin{itemize}
                \item Check security of third-party software libraries
                \item Vulnerability testing
                \item ``Public key/private key hashing''
                \item Fuzzing tools
                \item Encryption tools
                \item More secure versions of existing tools (e.g., OpenLST)
                \item Documentation
                \begin{itemize}
                    \item Example code
                    \item Best practices descriptions
                \end{itemize}
            \end{itemize}
        \end{itemize}
    \end{itemize}
\end{itemize}
\section{Repository Information}
\label{appendix:repositories}

\subsection{Repository Statistics per Satellite Club}
\label{appendix:repositories:repo-stats}

Tables \ref{tbl:repos:satelliteA}, \ref{tbl:repos:satelliteB}, and \ref{tbl:repos:satelliteC} show repository statistics for each satellite club where we performed a code analysis. For each group, we manually categorize repositories based on both the subsystem they are for and the type of design collateral they contain. We collect this information by first cloning each repository from the satellite clubs' GitHub organizations. Then, we manually categorize each repository and parse the Git histories to extract author/commit information. Note that totals may not sum up across the rows, since repositories may address multiple subsystems or design collateral types.

Entries marked with asterisks (*) represent outliers which we can explain in the data:

\begin{itemize}
    \item Table~\ref{tbl:repos:satelliteA} (miscellaneous commits): this satellite club used an automated script to update some of their repositories, hence the high commit count.
    \item Table~\ref{tbl:repos:satelliteC} (flight computer authors): this satellite club forked an industry-standard boot firmware into their organization. Our Git author counts also include all (non-satellite club) committers who contributed to the fork's upstream repository.
\end{itemize}

\begin{table*}[ht!]
\centering
\begin{tabular}{l|l l l|l l l|l l l|l l l|l l l}
    \textit{\satelliteA} &  \multicolumn{3}{c|}{Hardware} & \multicolumn{3}{c|}{Software} & \multicolumn{3}{c|}{Infrastructure} & \multicolumn{3}{c|}{Documentation} & \multicolumn{3}{c}{\textbf{Total}} \\
                      & R & A & C & R & A & C & R & A & C & R & A & C & R & A & C \\
    \hline
Ground Station & 0 &0& 0 &11 &35& 786 &0 &0& 0 &2 &7& 280 & 13 &35& 1066 \\
Sensors & 0 &0& 0 &5 &10& 63 &1 &2& 4 &0 &0& 0 & 6 &10& 67 \\
Communications & 1 &3& 24 &5 &13& 217 &1 &1& 8 &0 &0& 0 & 7 &16& 249 \\
Bus & 0 &0& 0 &0 &0& 0 &0 &0& 0 &0 &0& 0 & 0 &0& 0 \\
Power & 3 &3& 9 &0 &0& 0 &0 &0& 0 &0 &0& 0 & 3 &3& 9 \\
Flight Computer & 1 &14& 256 &9 &38& 1145 &1 &1& 2 &0 &0& 0 & 11 &51& 1403 \\
Actuators & 0 &0& 0 &1 &4& 39 &0 &0& 0 &0 &0& 0 & 1 &4& 39 \\
Educational & 0 &0& 0 &3 &9& 62 &0 &0& 0 &2 &3& 67 & 5 &10& 129 \\
Website & 0 &0& 0 &8 &32& 330 &2 &9& 34 &2 &6& 24 & 12 &39& 388 \\
Miscellaneous & 0 &0& 0 &4 &146& 6339* &3 &115& 5782* &0 &0& 0 & 7 &146& 12121 \\
\hline
\textbf{Total} & 5 &19& 289 &40 &244& 8732 &8 &125& 5830 &6 &15& 371 & 53 &270& 9135 \\
\end{tabular}
\caption{Number of \textbf{R}epositories, unique \textbf{A}uthors, and \textbf{C}ommits per subsystem and type for \satelliteA.}
\label{tbl:repos:satelliteA}
\end{table*}

\begin{table*}[ht!]
\centering
\begin{tabular}{l|l l l|l l l|l l l|l l l|l l l}
    \textit{\satelliteB} &   \multicolumn{3}{c|}{Hardware} & \multicolumn{3}{c|}{Software} & \multicolumn{3}{c|}{Infrastructure} & \multicolumn{3}{c|}{Documentation} & \multicolumn{3}{c}{\textbf{Total}} \\
                          & R & A & C & R & A & C & R & A & C & R & A & C & R & A & C \\
    \hline
Ground Station & 0 &0& 0 &1 &13& 462 &0 &0& 0 &0 &0& 0 & 1 &13& 462 \\
Sensors & 4 &4& 62 &7 &32& 521 &1 &1& 31 &0 &0& 0 & 12 &35& 614 \\
Communications & 1 &1& 11 &2 &5& 27 &0 &0& 0 &0 &0& 0 & 3 &5& 38 \\
Bus & 0 &0& 0 &0 &0& 0 &0 &0& 0 &0 &0& 0 & 0 &0& 0 \\
Power & 5 &6& 79 &2 &5& 45 &1 &1& 11 &0 &0& 0 & 8 &7& 135 \\
Flight Computer & 3 &6& 54 &5 &40& 2199 &1 &1& 3 &1 &37& 2138 & 10 &44& 4394 \\
Actuators & 8 &11& 75 &7 &10& 131 &4 &10& 62 &0 &0& 0 & 19 &18& 268 \\
Educational & 1 &8& 12 &3 &10& 49 &0 &0& 0 &3 &7& 20 & 7 &20& 81 \\
Website & 20 &0& 0 &0 &0& 0 &2 &2& 9035 &1 &7& 78 & 3 &9& 9113 \\
Miscellaneous & 3 &8& 62 &4 &22& 546 &2 &12& 63 &0 &0& 0 & 9 &29& 671 \\
\hline
\textbf{Total} & 24 &22& 328 &30 &88& 3518 &11 &22& 9205 &5 &47& 2236 & 60 &107& 12916 \\
\end{tabular}
\caption{Number of \textbf{R}epositories, unique \textbf{A}uthors, and \textbf{C}ommits per subsystem and type for \satelliteB.}
\label{tbl:repos:satelliteB}
\end{table*}

\begin{table*}[ht!]
\centering
\begin{tabular}{l|l l l|l l l|l l l|l l l|l l l}
    \textit{\satelliteC} & \multicolumn{3}{c|}{Hardware} & \multicolumn{3}{c|}{Software} & \multicolumn{3}{c|}{Infrastructure} & \multicolumn{3}{c|}{Documentation} & \multicolumn{3}{c}{\textbf{Total}} \\
                          & R & A & C & R & A & C & R & A & C & R & A & C & R & A & C \\
    \hline
Ground Station & 1 &19& 282 &2 &19& 292 &0 &0& 0 &0 &0& 0 & 3 &19& 574 \\
Sensors & 6 &43& 920 &16 &73& 3285 &3 &22& 1848 &2 &4& 10 & 27 &99& 6063 \\
Communications & 5 &35& 1266 &10 &50& 1255 &1 &2& 2 &0 &0& 0 & 16 &72& 2523 \\
Bus & 9 &71& 2443 &3 &15& 573 &6 &49& 668 &1 &2& 7 & 19 &83& 3691 \\
Power & 5 &30& 817 &2 &9& 167 &4 &17& 355 &0 &0& 0 & 11 &30& 1339 \\
Flight Computer & 4 &36& 930 &10 &3444*& 97028* &5 &31& 1986 &1 &2& 7 & 20 &3463& 99951 \\
Actuators & 2 &11& 109 &1 &17& 1757 &1 &17& 1757 &1 &2& 3 & 5 &29& 3626 \\
Educational & 0 &0& 0 &0 &0& 0 &0 &0& 0 &5 &26& 175 & 5 &26& 175 \\
Website & 0 &0& 0 &4 &14& 128 &1 &2& 10 &1 &2& 5 & 6 &15& 143 \\
Miscellaneous & 2 &33& 617 &6 &19& 368 &4 &13& 317 &1 &1& 2 & 13 &52& 1304 \\
\hline
\textbf{Total} & 32 &157& 7172 &50 &3571& 102745 &22 &100& 5172 &9 &29& 192 & 90 &3664& 109584 \\
\end{tabular}
\caption{Number of \textbf{R}epositories, unique \textbf{A}uthors, and \textbf{C}ommits per subsystem and type for \satelliteC.}
\label{tbl:repos:satelliteC}
\end{table*}

\subsection{Number of Files per Software Type}

We now focus on only the repositories we categorized as containing "software". Tables \ref{tbl:files:satelliteA}, \ref{tbl:files:satelliteB}, and \ref{tbl:files:satelliteC} show the number of source code files for each of the satellite club's most commonly used languages. This data was collected by using the commonly used \texttt{cloc}~\footnote{\url{https://linux.die.net/man/1/cloc}} utility, which counts the number of source lines for a large array of recognizable programming languages.

For all satellite clubs, but especially \satelliteA, we see a high number of C/C++ headers. This is because these satellite clubs vendored various SDKs for embedded devices directly into their repositories. \satelliteA also implemented a mobile app, the repository for which includes vendored Objective-C libraries.

\begin{table*}[ht!]
\centering
\begin{tabular}{l|c c c c c c c|c}
    \textit{\satelliteA} & C & C++ & C/C++ Header* & Java & Javascript & Objective-C* & Python & \textbf{Total}  \\
    \hline
    Ground Station  & 16    & 230   & 13472 & 148   & 120   & 884   & 610   & 15480  \\
    Sensors         & 60    & 0     & 264   & 0     & 0     & 0     & 6     & 330 \\
    Communications  & 0     & 0     & 0     & 0     & 131   & 0     & 24    & 155 \\
    Bus             & 0     & 0     & 0     & 0     & 0     & 0     & 0     & 0 \\
    Power           & 0     & 0     & 0     & 0     & 0     & 0     & 0     & 0\\
    Flight Computer & 268   & 0     & 1115  & 0     & 0     & 0     & 4     & 1387 \\
    Actuators       & 0     & 0     & 0     & 0     & 0     & 6     & 0     & 6 \\
    Educational     & 35    & 0     & 199   & 0     & 0     & 0     & 0     & 234 \\
    Website         & 3     & 211   & 13427 & 148   & 547   & 884   & 0     & 15220 \\
    Miscellaneous   & 52    & 32    & 250   & 0     & 77    & 0     & 0     & 411 \\
    \hline
    \textbf{Total}  & 330   & 262   & 14786 & 148   & 757   & 884   & 644   & 17811 \\
\end{tabular}
\caption{Number of source code files per subsystem and language for \satelliteA.}
\label{tbl:files:satelliteA}
\end{table*}

\begin{table*}[ht!]
\centering
\begin{tabular}{l|c c c c c c|c}
    \textit{\satelliteB}    & Arduino Sketch    & C     & C++   & C/C++ Header  & Python    & Ruby  & \textbf{Total}  \\
    \hline
    Ground Station      & 0                 & 0     & 0     & 0             & 8         & 88    & 96 \\
    Sensors             & 3                 & 27    & 49    & 113           & 20        & 24    & 236 \\
    Communications      & 0                 & 0     & 0     & 0             & 35        & 0     & 35 \\
    Bus                 & 0                 & 0     & 0     & 0             & 0         & 0     & 0 \\
    Power               & 6                 & 0     & 0     & 0             & 0         & 0     & 6 \\
    Flight Computer     & 6                 & 216   & 16    & 275           & 21        & 216   & 750 \\
    Actuators           & 1                 & 5     & 11    & 21            & 3         & 0     & 41 \\
    Educational         & 0                 & 4     & 7     & 8             & 0         & 0     & 19 \\
    Website             & 0                 & 0     & 0     & 0             & 0         & 0     & 0 \\
    Miscellaneous       & 2                 & 2     & 5     & 14            & 8         & 88    & 119 \\
    \hline
    \textbf{Total}      & 18                & 254   & 88    & 431           & 87        & 275   & 1153\\
\end{tabular}
\caption{Number of source code files per subsystem and language for \satelliteB.}
\label{tbl:files:satelliteB}
\end{table*}

\begin{table*}[ht!]
\centering
\begin{tabular}{l|c c c c c c|c}
    \textit{\satelliteC}& Assembly  & C     & C++   & C/C++ Header  & Python    & Rust  & \textbf{Total}  \\
    \hline
    Ground Station  & 0         & 22    & 0     & 300           & 2         & 0     & 324 \\
    Sensors         & 274       & 554   & 20    & 1076          & 110       & 2*     & 2036 \\
    Communications  & 0         & 130   & 3     & 395           & 22        & 18    & 568 \\
    Bus             & 0         & 25    & 0     & 18            & 68        & 0     & 111 \\
    Power           & 0         & 0     & 3     & 4             & 10        & 0     & 17\\
    Flight Computer & 0         & 48    & 0     & 45            & 108       & 3     & 204 \\
    Actuators       & 0         & 132   & 2     & 176           & 5         & 0     & 315\\
    Educational     & 0         & 0     & 0     & 0             & 0         & 0     & 0\\
    Website         & 0         & 0     & 0     & 0             & 0         & 0     & 0 \\
    Miscellaneous   & 1         & 1     & 0     & 0             & 36        & 0     & 38\\
    \hline
    \textbf{Total}  & 275       & 728   & 26    & 1217          & 311       & 23    & 2580 \\
\end{tabular}
\caption{Number of source code files per subsystem and language for \satelliteC.}
\label{tbl:files:satelliteC}
\end{table*}

\subsection{Number of Code Lines per Software Type}

Tables \ref{tbl:code:satelliteA}, \ref{tbl:code:satelliteB}, and \ref{tbl:code:satelliteC} use the same underlying data from \texttt{cloc}, but instead display the number of lines characterized as "source code" (as opposed to "blank lines" or "comments"). Again, C/C++ headers from vendored SDKs comprise a large proportion of the overall codebase. \satelliteC also includes two huge Rust files which appear to be automatically generated libraries for interfacing with specific microcontrollers.

\begin{table*}[ht!]
\centering
\begin{tabular}{l|c c c c c c c|c}
    \textit{\satelliteA} & C & C++ & C/C++ Header & Java & Javascript & Objective-C & Python & \textbf{Total}  \\
    \hline
    Ground Station  & 5850  & 49144 & 1703803   & 33883 & 7455  & 113530    & 142245    & 2055910 \\
    Sensors         & 12532 & 0     & 128926    & 0     & 0     & 0         & 771       & 142229 \\
    Communications  & 0     & 0     & 0         & 0     & 6014  & 0         & 1857      & 7871 \\
    Bus             & 0     & 0     & 0         & 0     & 0     & 0         & 0         & 0 \\
    Power           & 0     & 0     & 0         & 0     & 0     & 0         & 0         & 0 \\
    Flight Computer & 59954 & 0     & 448973    & 0     & 0     & 0         & 245       & 509172 \\
    Actuators       & 0     & 0     & 0         & 0     & 0     & 771       & 0         & 771 \\
    Educational     & 3630  & 0     & 108735    & 0     & 0     & 0         & 0         & 112365 \\
    Website         & 4773  & 47687 & 1702967   & 33883 & 110853& 113530    & 0         & 2013693 \\
    Miscellaneous   & 4539  & 6591  & 114756    & 0     & 151790& 0         & 0         & 277676 \\
    \hline
    \textbf{Total}  & 71825 & 55735 & 2209085   & 33883 & 268059& 113530    & 145118    & 2897235\\
\end{tabular}
\caption{Number of source code lines per subsystem and language for \satelliteA.}
\label{tbl:code:satelliteA}
\end{table*}

\begin{table*}[ht!]
\centering
\begin{tabular}{l|c c c c c c|c}
    \textit{\satelliteB}    & Arduino Sketch    & C     & C++   & C/C++ Header  & Python    & Ruby  & \textbf{Total}  \\
    \hline
    Ground Station      & 0                 & 0     & 0     & 0             & 1645      & 1776  & 3421 \\
    Sensors             & 203               & 15640 & 8045  & 60076         & 1069      & 264   & 85297 \\
    Communications      & 0                 & 0     & 0     & 0             & 3515      & 0     & 3515 \\
    Bus                 & 0                 & 0     & 0     & 0             & 0         & 0     & 0 \\
    Power               & 325               & 0     & 0     & 0             & 0         & 0     & 325 \\
    Flight Computer     & 700               & 45408 & 10893 & 44224         & 1224      & 11888 & 114337\\
    Actuators           & 60                & 1142  & 568   & 971           & 114       & 0     & 2855 \\
    Educational         & 0                 & 209   & 174   & 203           & 0         & 0     & 586 \\
    Website             & 0                 & 0     & 0     & 0             & 0         & 0     & 0 \\
    Miscellaneous       & 63                & 274   & 114   & 14429         & 1645      & 1776  & 18301 \\
    \hline
    \textbf{Total}      & 1351              & 62673 & 19794 & 119903        & 7567      & 13522 & 224810\\
\end{tabular}
\caption{Number of source code lines per subsystem and language for \satelliteB.}
\label{tbl:code:satelliteB}
\end{table*}

\begin{table*}[ht!]
\centering
\begin{tabular}{l|c c c c c c|c}
    \textit{\satelliteC}& Assembly  & C     & C++   & C/C++ Header  & Python    & Rust  & \textbf{Total}  \\
    \hline
    Ground Station  & 0         & 6018  & 0     & 72450         & 80        & 0     & 78548 \\
    Sensors         & 85802     & 234261& 3633  & 1059330       & 8804      & 449665*& 1841495 \\
    Communications  & 0         & 37138 & 268   & 77353         & 1120      & 8145  & 124024 \\
    Bus             & 0         & 4016  & 0     & 442           & 5113      & 0     & 9571 \\
    Power           & 0         & 0     & 515   & 238           & 1146      & 0     & 1899\\
    Flight Computer & 0         & 5741  & 0     & 7338          & 10239     & 364   & 23682 \\
    Actuators       & 0         & 26041 & 43    & 45936         & 509       & 0     & 72529\\
    Educational     & 0         & 0     & 0     & 0             & 0         & 0     & 0\\
    Website         & 0         & 0     & 0     & 0             & 0         & 0     & 0 \\
    Miscellaneous   & 140       & 6     & 0     & 0             & 3591      & 0     & 3737\\
    \hline
    \textbf{Total}  & 85942     & 274719& 4416  & 1068598       & 26935     & 458174& 1918784 \\
\end{tabular}
\caption{Number of source code lines per subsystem and language for \satelliteC.}
\label{tbl:code:satelliteC}
\end{table*}

\end{document}